\documentclass[preprint,showpacs,amsmath,amssymb]{revtex4}
\usepackage{graphicx,amsmath,amssymb,bbm,times}
\begin{document}

\title{Self-consistent characterization of light statistics}

\author{Maria Bondani} \email{maria.bondani@uninsubria.it}
\affiliation{National Laboratory for Ultrafast and Ultraintense
Optical Science - C.N.R.-I.N.F.M.\\ and C.N.I.S.M., U.d.R. Como, I-22100, Como, Italy}
\author{Alessia Allevi}
\affiliation{C.N.I.S.M., U.d.R. Como, I-22100, Como, Italy}
\author{Andrea Agliati}
\affiliation{Dipartimento di Fisica e Matematica, Universit\`a degli Studi dell'Insubria, I-22100, Como, Italy}
\author{Alessandra Andreoni}
\affiliation{Dipartimento di Fisica e
Matematica, Universit\`a degli Studi dell'Insubria\\ and C.N.I.S.M., U.d.R. Como, I-22100, Como, Italy}

\begin{abstract}

We demonstrate the possibility of a self-consistent characterization of the photon-number statistics of a light field by using photoemissive detectors with internal gain simply endowed with linear input/output responses. The method can be applied to both microscopic and mesoscopic photon-number regimes. The detectors must operate in the linear range without need of photon-counting capabilities.\bigskip
\end{abstract}
\maketitle

\section{Introduction}

Although the measurement of photon-number statistics gives an
incomplete description of the optical state, it can be helpful to provide fundamental information on the nature of any optical field and hence to
discriminate between light of different kinds, either in the
classical or non-classical domain.

Measuring the photon-number statistics is a difficult task as it requires a complete knowledge of the entire detection process.
In some situations, the evaluation of the first two statistical
momenta can be enough for characterizing light statistics. In particular, mean value and variance, combined into the Fano factor, can discriminate between different statistics, even if sometimes the evaluation of higher order momenta is necessary.

In this paper we demonstrate that it is possible to
determine the photon-number statistics by means of a direct
intensity measurement performed with linear photodetectors endowed
with internal gain. In order to bypass the problem of calibrating the detector, we devise a  measurement technique that takes advantage of the
linearity properties of the detector and is based on the measurement
of the same state at different mean numbers of photons. In this way we simultaneously obtain the calibration of the detection chain and the reconstruction of the photoelectron statistics.

Our technique has the additional advantage of being applicable to fields in the mesoscopic intensity regime (less than 1000 photons), which is scarcely explored.

The competing techniques for determining the photon-number
distributions are for instance: tomography with homodyne detection \cite{munroe1995,zhang2002,raymer2004}, measurement with internal-gain photodetectors able to resolve peaks of different photoelectron numbers, plus analysis of peak integrals in order to
reconstruct the photoelectron number distribution \cite{zambra2004}. Drawbacks of the latter approach are: the low-limit in the mean number of detected photons due to the low quantum
efficiency; the need of independent calibration of the detector; the need of a guess on the statistics to which the data must be fitted, as there is no direct indication from the
measurement \cite{zambra2004}. If the measurements are performed with linear photodiodes lacking internal gain, the drawbacks is the low-limit to the mean number of detected photons (p-i-n photodiodes can detect down to one thousand
photons upon external amplification \cite{paleari2004,bondani2007}). A further alternative is represented by ON/OFF measurements combined with maximum likelihood reconstruction \cite{zambra2005}.
Section \ref{sec:teo} presents the theoretical model we will use to interpret the experimental data, Section \ref{sec:exp} describes the experimental setup used for the measurements and Section \ref{sec:analysis} presents the technique for analyzing the data and the experimental results. Finally we draw some conclusions.

\section{Theory}\label{sec:teo}

\begin{figure}
\begin{center}
\resizebox{0.6\columnwidth}{!}{%
 \includegraphics[width=0.4\textwidth,angle=270]{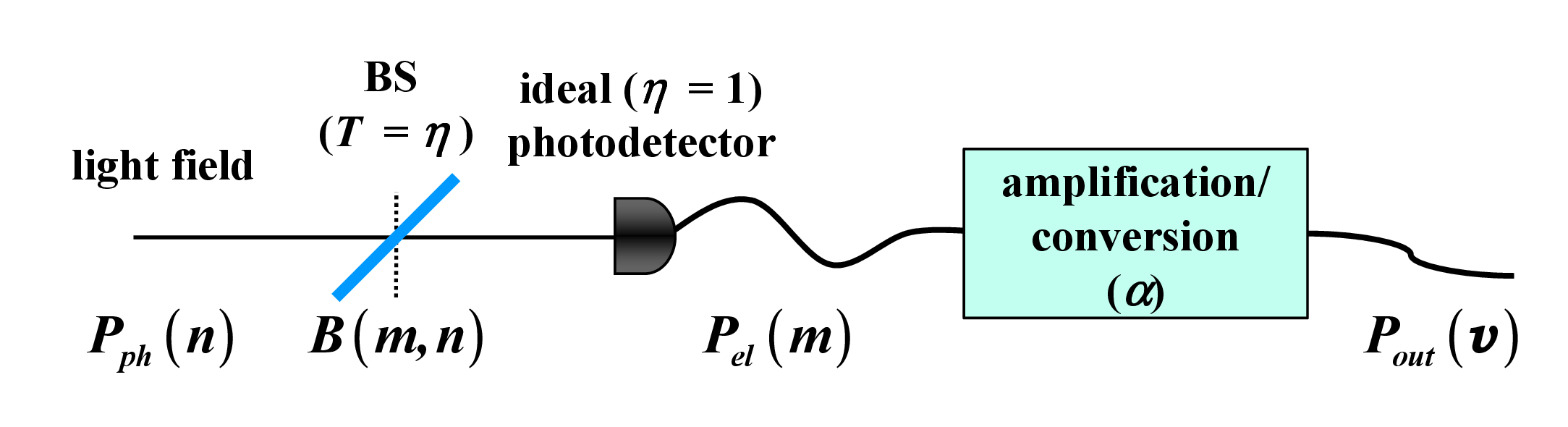} }
\caption{Sketch of the theoretical equivalent of a photodetector with $\eta<1$ endowed with internal gain, whose output signal is also externally processed to achieve an overall amplification by a factor $\alpha$.} \label{f:scheme}
\end{center}
\end{figure}

We can model the detection process as a two step process: photodetection
by the photocathode and amplification (both internal to the detector and external) (see Fig.~\ref{f:scheme}).

We start from the observation that a real photodetector, having quantum efficiency $\eta<1$, behaves like a perfect detector ($\eta=1$) preceded by a beam spitter (BS) with transmittance $T=\eta$ \cite{agliati2005,zambra2005}. The photodetection process can thus be described as the convolution of the photon statistics with the Bernoullian action of the BS \cite{mandel1995,agliati2005}.
The statistics of the number of
photoelectrons emitted by the photocathode, $P_{el}(m)$, is thus
linked to the statistics of the number of photons, $P_{ph}(n)$, by
\begin{equation}
 P_{el}(m)=\sum_{n=m}^{\infty} B(m,n) P_{ph}(n)=\sum_{n=m}^{\infty} \left(
 \begin{array}{c}n\\m\end{array}\right)
 \eta^m (1-\eta)^{n-m} P_{ph}(n)\ .\label{eq:phel}
\end{equation}
If we limit our analysis to the first two momenta of the
distributions, the links between the statistics of photons and photoelectrons are given by \cite{agliati2005}
\begin{equation}
 \bar{m} = \eta \bar{n}\ ;\  \sigma_{el}^2(m)= \eta^2\sigma_{ph}^2(n)+\eta(1-\eta)\bar{n}
\label{eq:momentsPHEL}\ ,
\end{equation}
where $\bar{n}=\sum_{n=0}^{\infty} P_{ph}(n) n$ is the mean value and $\sigma_{ph}^2(n)=\sum_{n=0}^{\infty} P_{ph}(n) (n-\bar{n})^2$ is the variance. The same notation is adopted for photoelectrons.

A complete
description of the amplification process requires the detailed
knowledge, for each particular detector, of the amplification mechanism. However for the present discussion we limit ourselves to those detectors and to those operating regimes in
which the amplification process can be simply considered as linear: we adopt a strongly simplifying approach in which the spread of the single photoelectron peak in the detector output is negligible as compared to its mean value. The relation linking the statistics of
photoelectrons to that of the voltage outputs, even at the end of the
electronic chain external to the detector, is then
\begin{equation}
 P_{out}(v) = \frac{1}{\alpha} P_{el}(\alpha m)\ ,\label{eq:outs}
\end{equation}
being $\alpha$ the amplification/conversion coefficient given by the overall amplification
process and signal-processing electronics. Quite obviously, Eq.~(\ref{eq:outs}) shows that in general the distribution of the output voltage values is different from that of the photoelectrons. For the first two momenta of the distributions, the experimental final outputs are given by
\cite{agliati2005}
\begin{eqnarray}
 \bar{v}=\alpha \bar{m}\ ;\ \sigma_{out}^2(v) &=& \alpha^2 \sigma_{el}^2(m)\ ,\label{eq:momentsVOUT}
\end{eqnarray}
where the symbols are defined as above.
\begin{figure}
\begin{center}
\resizebox{0.6\columnwidth}{!}{%
 \includegraphics[width=0.4\textwidth,angle=270]{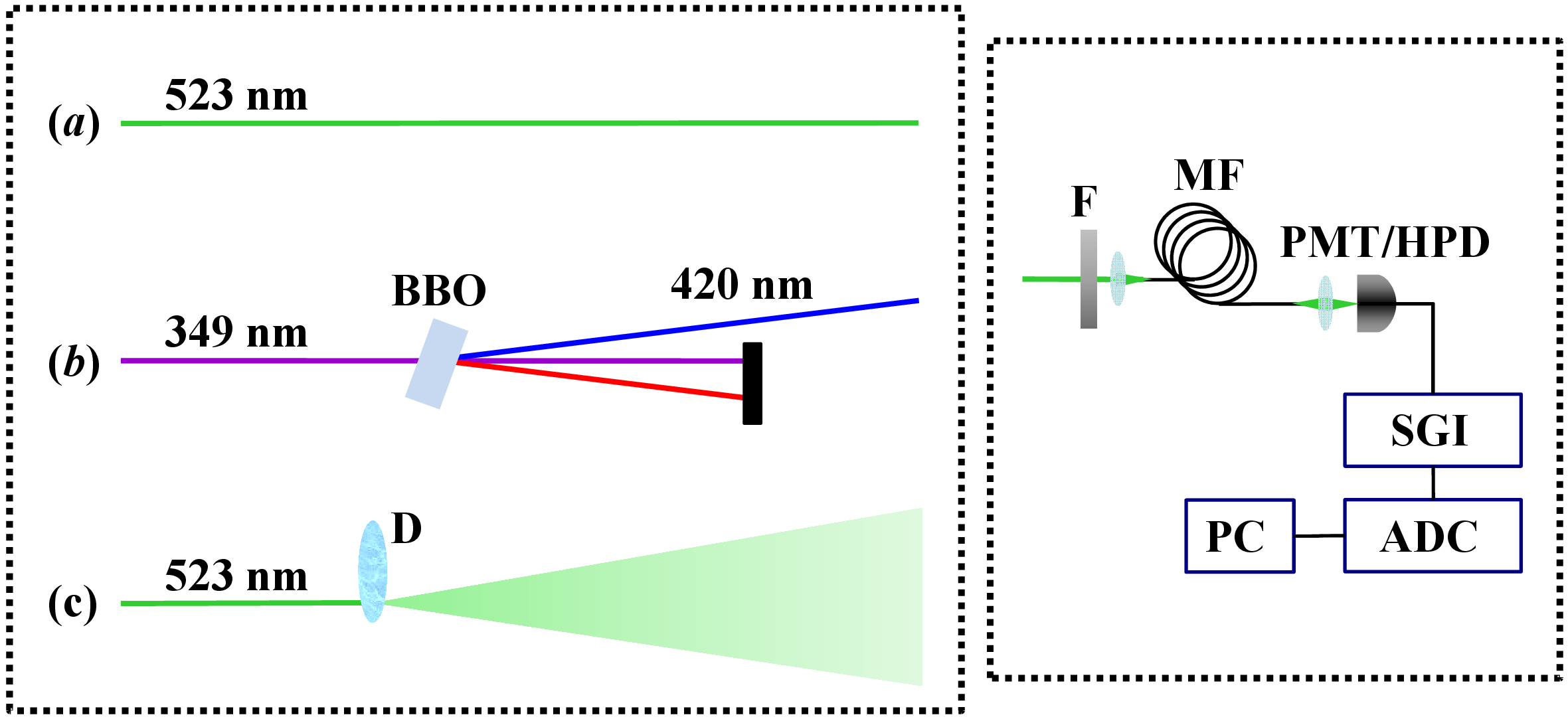} }
\caption{Sketch of the experimental setup. Left: light sources: (a) coherent state, (b) multimode thermal state and (c) multimode pseudo-thermal state. D, rotating ground-glass diffuser; BBO, type I nonlinear crystal. Right: measuring apparatus: F,
variable neutral density filter; MF, multimode fiber; PMT/HPD, detectors with internal gain;
SGI, synchronous gated integrator; ADC, analog-to-digital converter.} \label{f:setup}
\end{center}
\end{figure}

The most difficult task to be performed for reconstructing $P_{el}$ from $P_{out}$ is the experimental determination of the parameter $\alpha$ in Eq.~(\ref{eq:outs}). This task can be tackled by following a procedure based on repeated
measurements of the same state of light at different values of the quantum efficiency by inserting neutral filters in front of the detector. In particular we evaluate the following quantity

\begin{eqnarray}
 F_v  = \frac{\sigma_{out}^2(v)}{\bar{v}} = \frac{{\alpha ^2 \left[
 {\eta ^2 \sigma_{ph}^2(n)  + \eta \left( {1 - \eta } \right)\bar{n}} \right]}}{{\alpha \eta \bar{n}}} = \alpha \eta F + \alpha \left( {1 - \eta } \right)=\alpha \eta Q + \alpha ,
\label{eq:fano}
\end{eqnarray}
where $F=\sigma_{ph}^2(n)/\bar{n}$ is the Fano-factor that we have re-written as $F=1+Q$, $Q$ being the Mandel Q-factor \cite{mandel1979}. We observe that $F_v$ is linearly dependent on $\eta$ and that $\alpha$ can be obtained as the limit of $F_v$ at $\eta\rightarrow 0$.
By multiplying and dividing by $\bar{n}$ we get
\begin{eqnarray}
 F_v  = \frac{Q}{\bar{n}} {\bar{v}} + \alpha\; ,
\label{eq:fanoQ}
\end{eqnarray}
which now expresses $F_v$ as a function of the mean value of the measured output voltage. Note that $\bar{v}$ varies because of the variation of $\eta$ and not because of the change in the measured state ($Q/\bar{n}$ remains the same). The expression in Eq.~(\ref{eq:fanoQ}) is general and all the information about the specific state under measurement is contained in the angular coefficient $Q/\bar{n}=(\sigma_{ph}^2(n)-\bar{n})/\bar{n}^2$, which is zero for Poissonian light, positive for classical super-Poissonian light and negative for nonclassical sub-Poissonian light. By measuring the same light state at different values of $\bar{v}$ and evaluating $F_v$, we can directly obtain the value of $\alpha$ and the ratio $Q/\bar{n}$. Of course, different light states may have, for some values of the parameters, the same value of $Q/\bar{n}$ and this prevents the possibility of the unique determination of the light statistics. Nevertheless, if an assumption can be made on the statistics of the measured light, the method allows one to check it immediately.

The expected expressions of $F_v$ for the states presented in this paper will be given in Section~\ref{sec:analysis}, where the results are discussed.
\begin{figure}
\begin{center}
\resizebox{0.6\columnwidth}{!}{%
 \includegraphics[width=0.4\textwidth,angle=270]{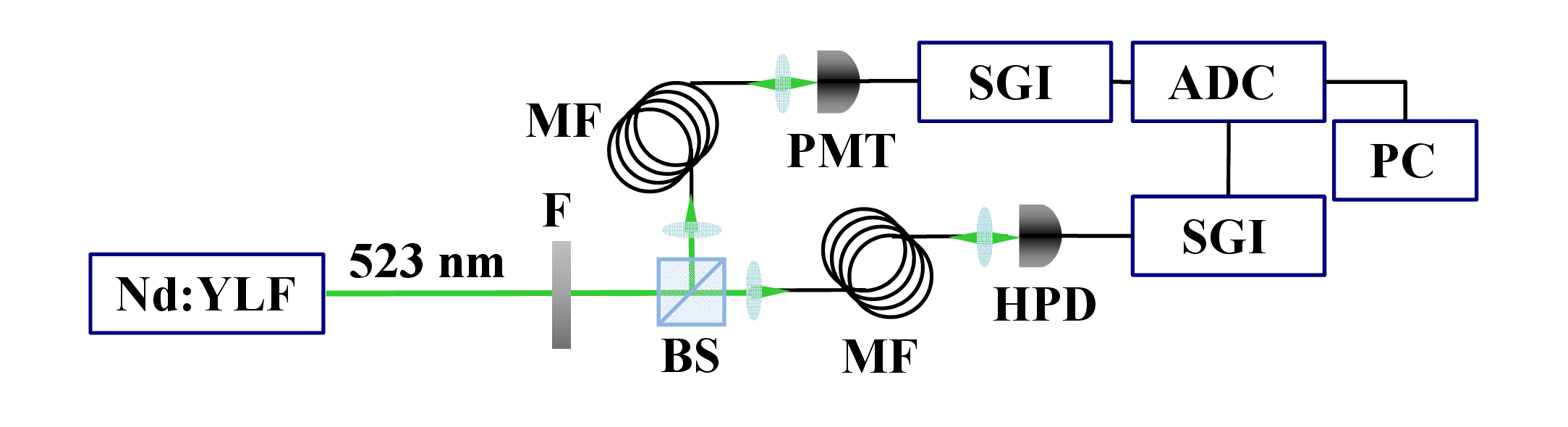} }
\caption{Sketch of the experimental setup for the comparison between PMT and HPD. Nd:YLF, laser source; BS, beam splitter; F, variable neutral density filter; MF, multimode fiber; PMT/HPD, detectors with internal gain; SGI, synchronous gated integrator; ADC, analog-to-digital converter.} \label{f:setupBIS}
\end{center}
\end{figure}

\section{Experiments}\label{sec:exp}

In Fig.~\ref{f:setup} we show the adopted experimental setup.
The light from the specific source to be investigated is collected by a multimode fiber (100 $\mu$m core diameter, OZ Optics, Canada). As the detector we used either a photomultiplier tube (PMT, 8850, Burle Industries, maximum quantum efficiency $\eta=0.24$ at 400~nm) or a hybrid photodiode module (HPD, H8236-40, Hamamatsu, Japan, maximum quantum efficiency $\eta=0.40$ at 550~nm). Both detectors are endowed with partial photon resolving capability and are linear over a wide range of intensities. The current outputs of the detectors are integrated (SGI, Stanford Research Systems), sampled and digitized (ADC, National Instruments). The final outputs are then recorded by a computer.

\begin{figure}
\begin{center}
\resizebox{0.6\columnwidth}{!}{%
 \includegraphics[width=0.4\textwidth,angle=270]{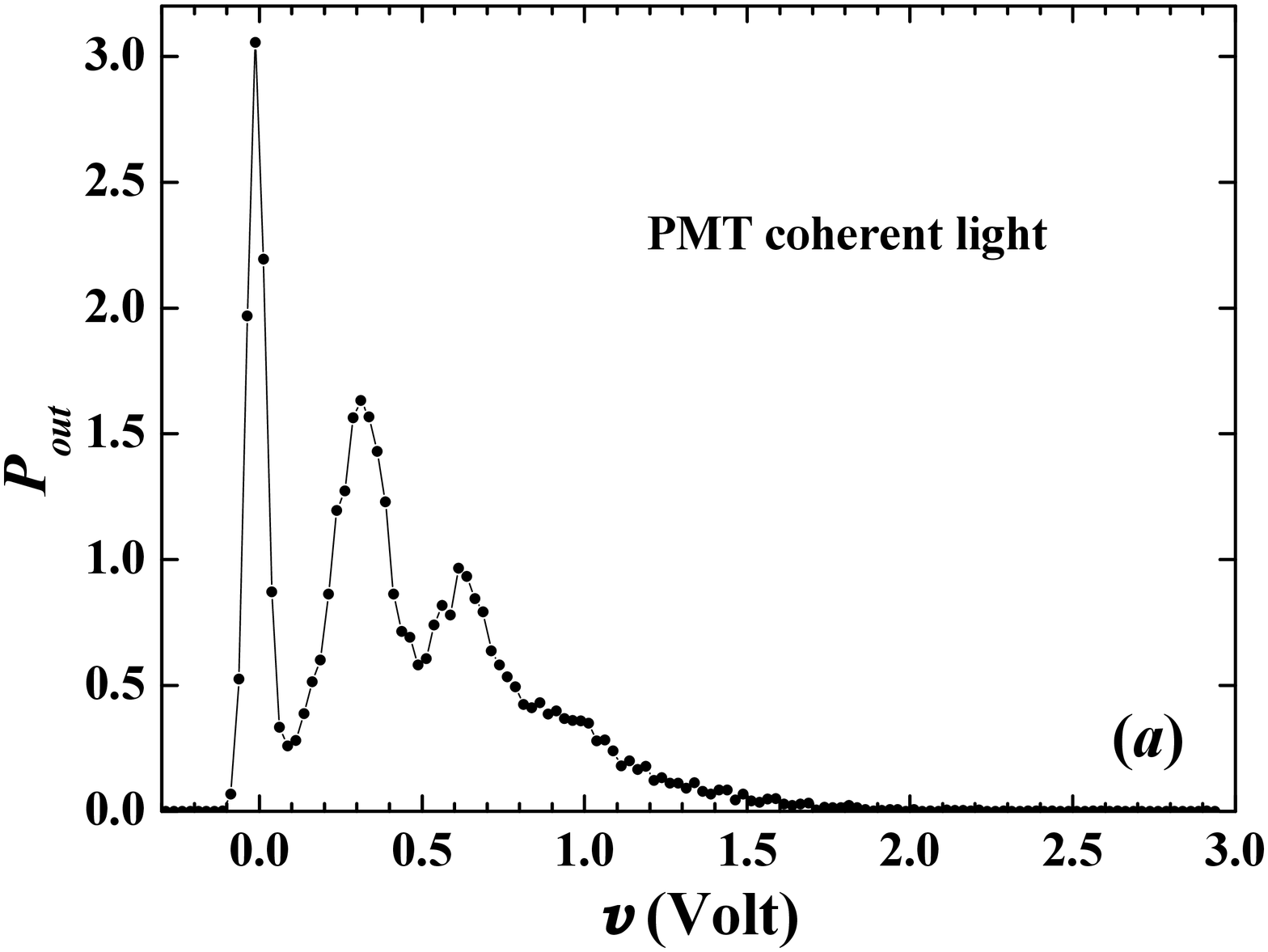}}
 \resizebox{0.6\columnwidth}{!}{%
 \includegraphics[width=0.4\textwidth,angle=270]{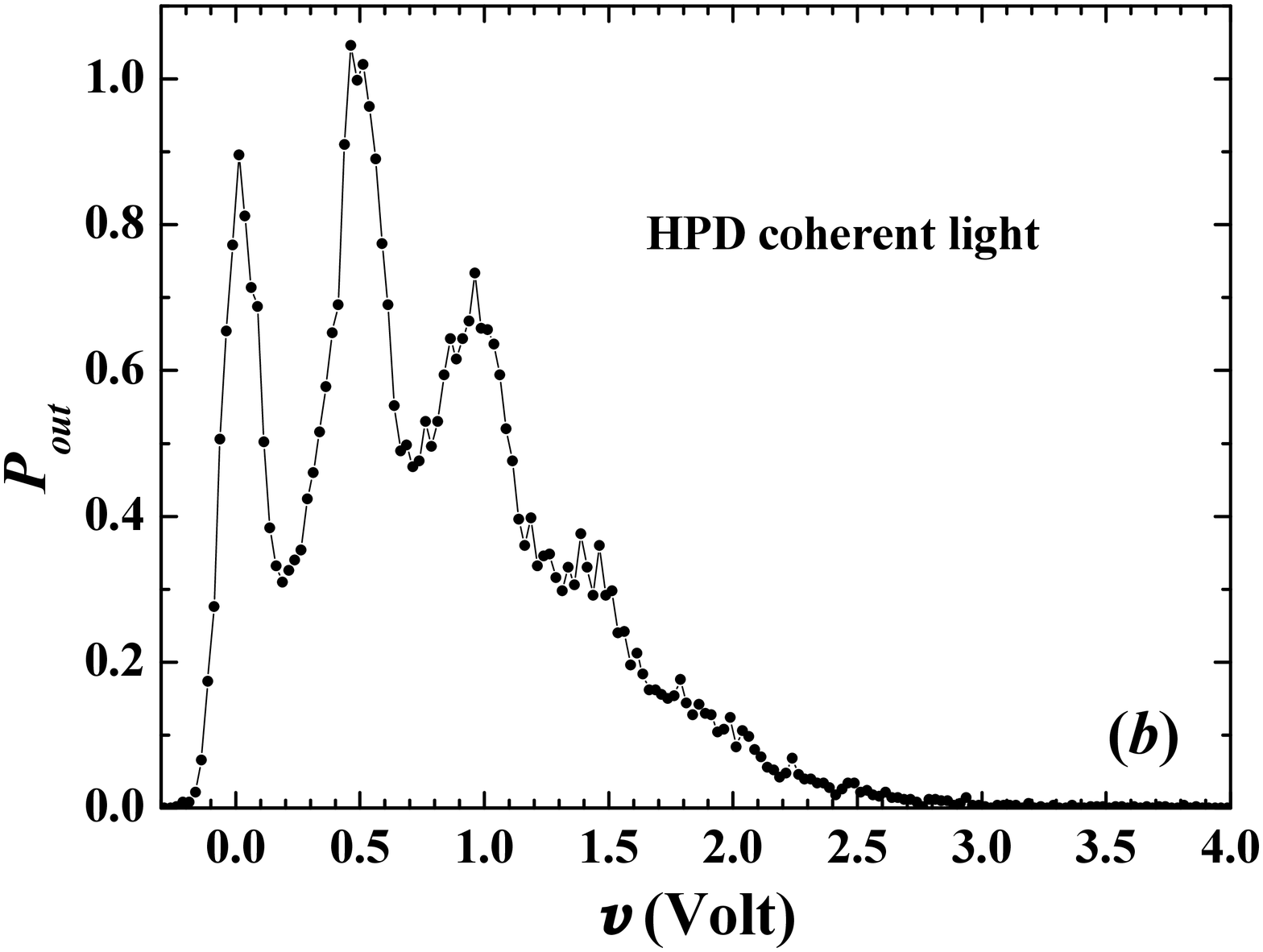}}
\caption{Typical pulse-height spectrum measured with the PMT ($a$) and with the HPD ($b$) simultaneously recorded at the outputs of a 50\% beam splitter, dividing a ps-pulsed coherent beam at 523~nm.} \label{f:response}
\end{center}
\end{figure}

As a preliminary check of the performances of the detectors, we simultaneously measured the light at the two outputs of a 50\% beam splitter (BS in Fig.~\ref{f:setupBIS}) dividing a ps-pulsed coherent beam at 523~nm. The light was obtained from the second harmonics of a Nd:YLF continuous-wave mode-locked laser amplified at 500~Hz repetition rate (High-Q Laser Production). In Fig.~\ref{f:response} we show a typical pulse-height spectrum of the PMT ($a$) and of the HPD ($b$). The zero value of the output voltage $v$ is set at the mean value of the response to dark of each detector, which was measured independently (see \textit{e.g.} the horizontal scale in Figs.~\ref{f:response} ($a$) and ($b$)).

A standard procedure for analyzing a pulse-height-spectrum is to find a suitable fit obtained as a multiple convolution of the function best fitting the single-photon peak \cite{zambra2004}. The procedure allows the reconstruction of a limited number of peaks (that depend on the resolving capabilities of the detectors) and fails in analyzing more intense fields when the pulse-height spectrum does not show well recognizable peaks.

The pulse-height spectrum of the detectors lose the peak structure as soon as the field intensity becomes mesoscopic, still remaining proportional to the field intensity.

\begin{figure}
\begin{center}
\resizebox{0.6\columnwidth}{!}{%
 \includegraphics[width=0.4\textwidth,angle=270]{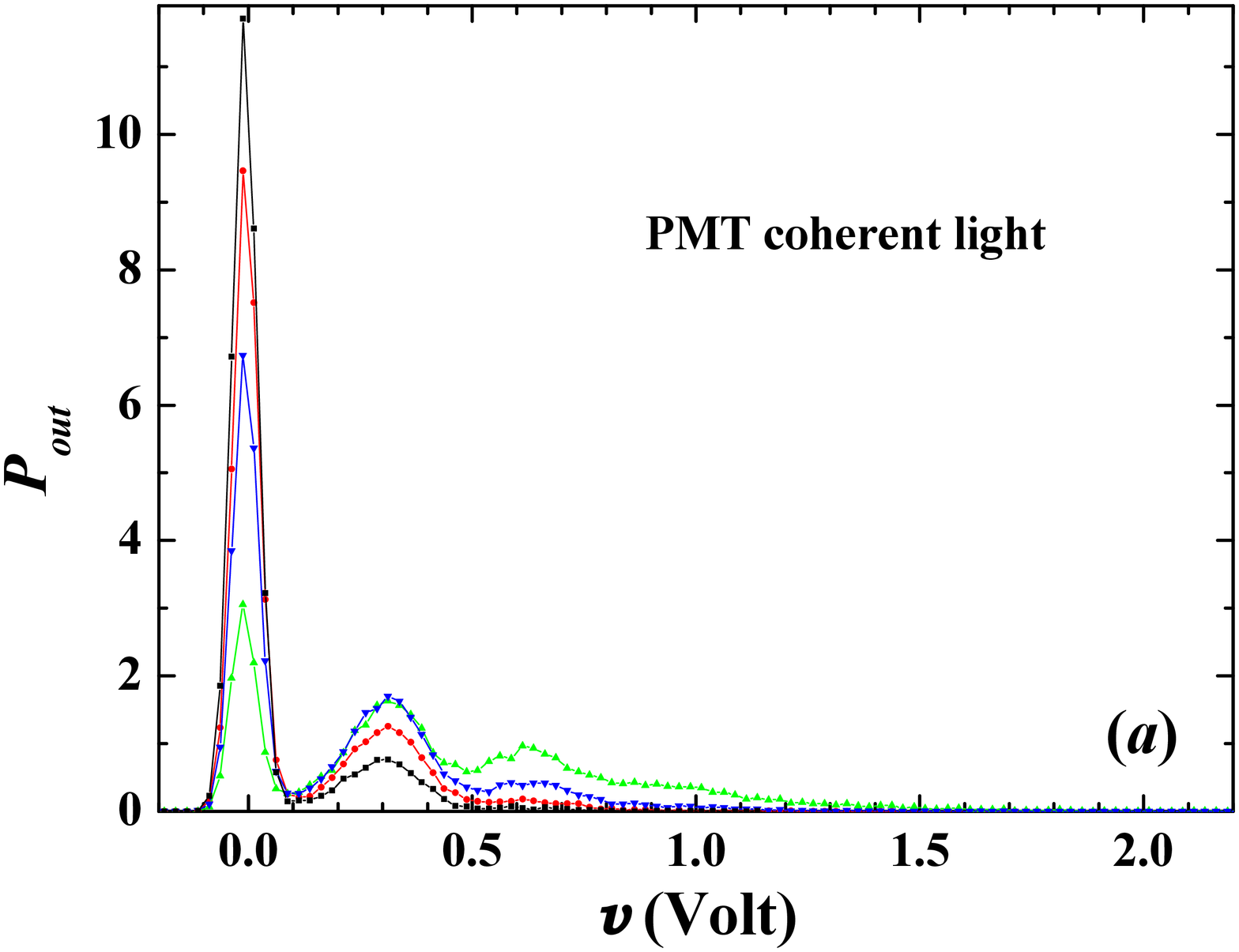}}
 \resizebox{0.6\columnwidth}{!}{%
 \includegraphics[width=0.4\textwidth,angle=270]{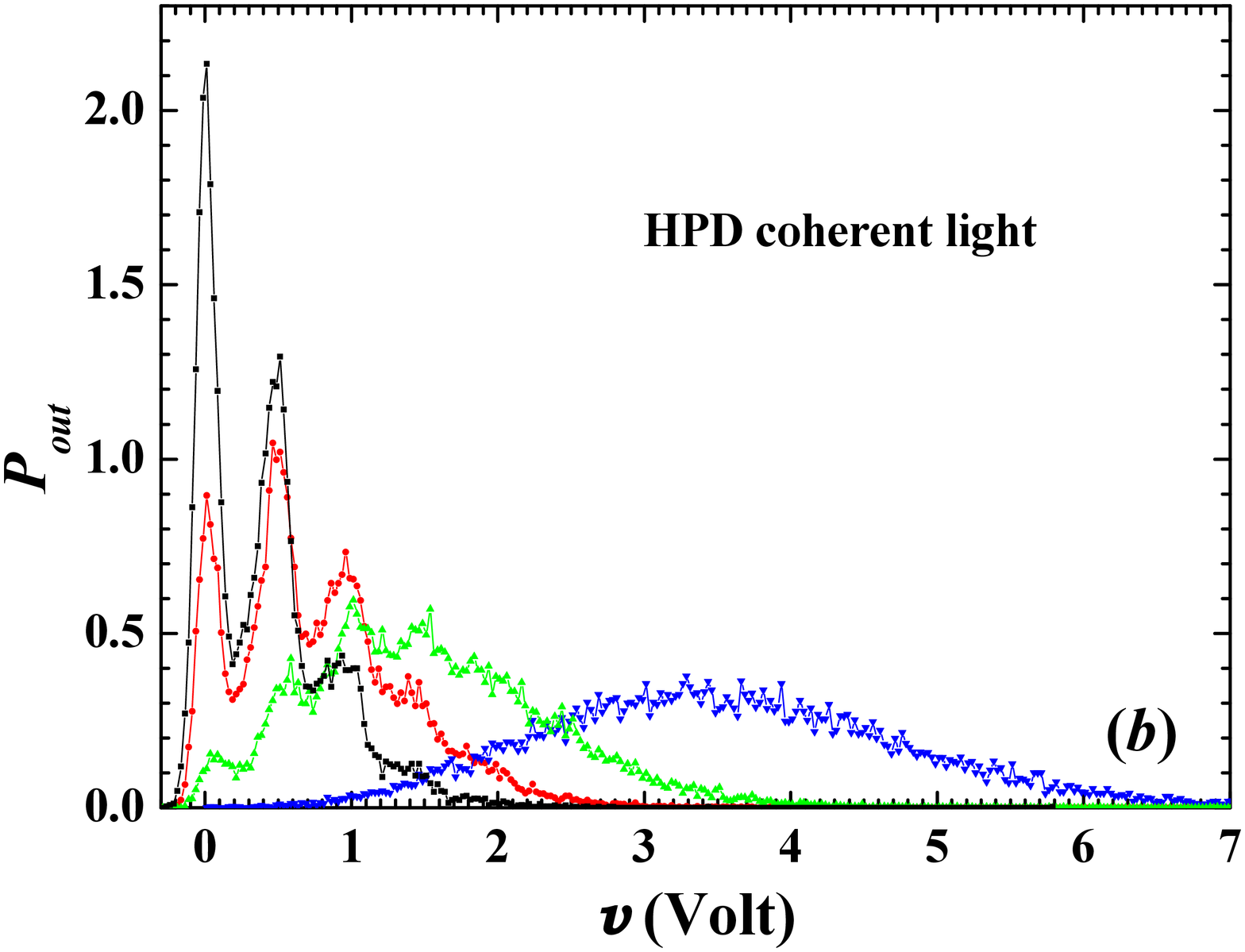}}
 \resizebox{0.6\columnwidth}{!}{%
 \includegraphics[width=0.4\textwidth,angle=270]{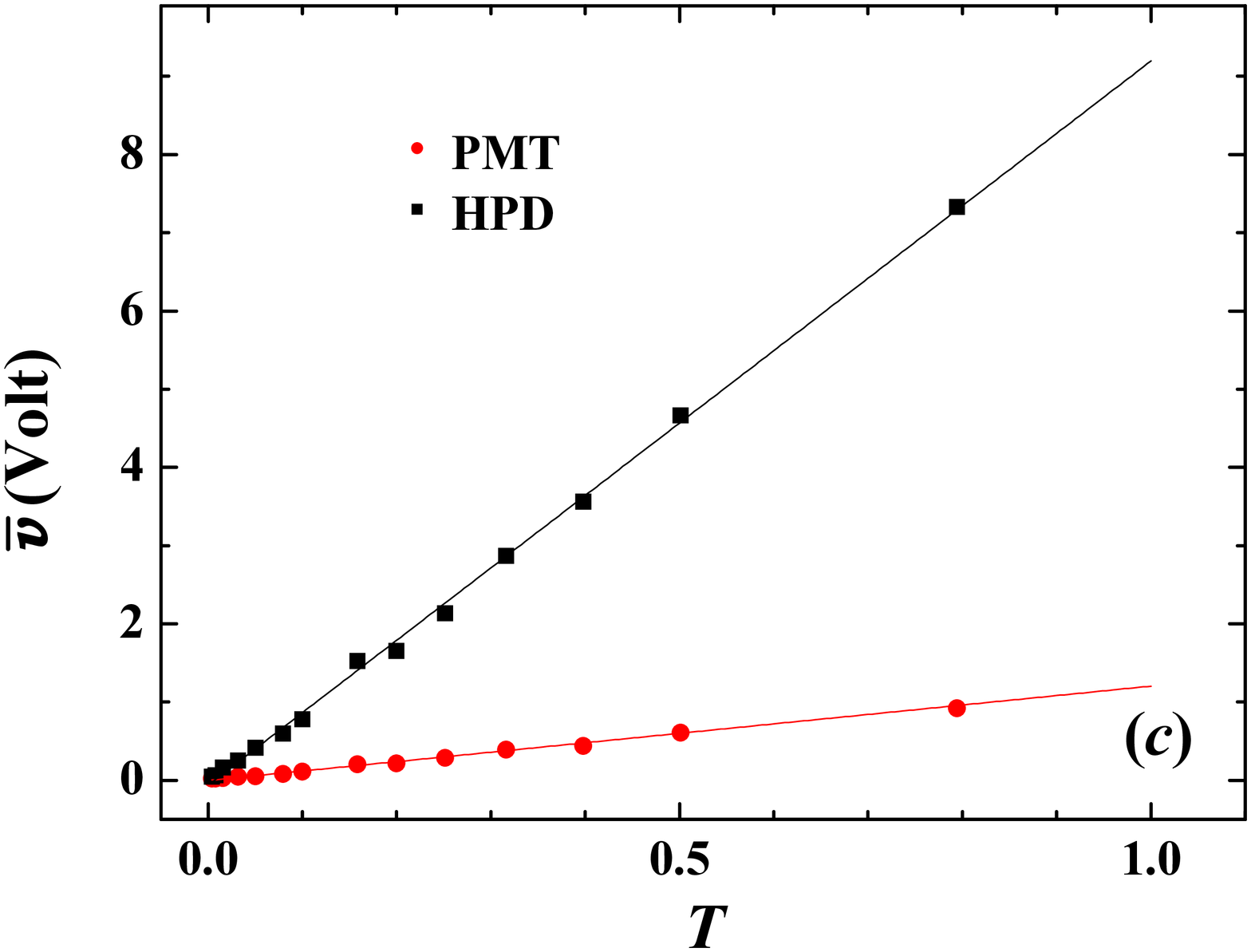}}
\caption{Pulse height spectra recorded by the PMT ($a$) and by the HPD ($b$) in the same configuration of Fig.~\ref{f:response} at different values of the transmittance of the neutral density filters. ($c$): Mean value of the output voltage as a function of the filters' transmittance. The difference in the angular coefficient reflect the different overall detection efficiency of the two arms.} \label{f:linearity}
\end{center}
\end{figure}

In Fig.~\ref{f:linearity} we show several pulse height spectra recorded by the PMT ($a$) and by the HPD ($b$) in the same configuration of Fig.~\ref{f:setupBIS}, in which neutral density filters are inserted on the beam path before the beam splitter. In Fig.~\ref{f:linearity} ($c$) we plot the mean values of the output voltage as a function of the transmittance of the filters: the difference between the two curves reflects the difference in the overall quantum efficiency of the two detectors at 523~nm. The linearity will play a key role in the analysis procedure described in Section~\ref{sec:analysis}.
\newpage
\section{Results and analysis}\label{sec:analysis}

We measured different light fields characterized by different statistics and at wavelengths that were chosen so as to match the highest quantum efficiencies of the detectors at best.
The aim of the measurements is to identify the type of light field, to determine the overall amplification/conversion coefficient $\alpha$ and to reconstruct the statistics of the photoelectrons. The steps of the experimental procedure are: 1) to measure the same field at different attenuations of the neutral density filters (the actual position of the filters depends on the specific measurement); 2) to evaluate the mean value and the variance of the output voltage; 3) to plot the experimental values of $F_v$ (see Eq.~(\ref{eq:fano})) as a function of the mean value $\bar{v}$; 4) to fit the plot to a straight line, thus obtain the value of the $\alpha$ factor, and to find, according to one of the theoretical models, the parameters of the photoelectron distribution; 5) to use the fitting parameters to recover the photoelectron statistics.

\begin{figure}
\begin{center}
\resizebox{0.6\columnwidth}{!}{%
 \includegraphics[width=0.4\textwidth,angle=270]{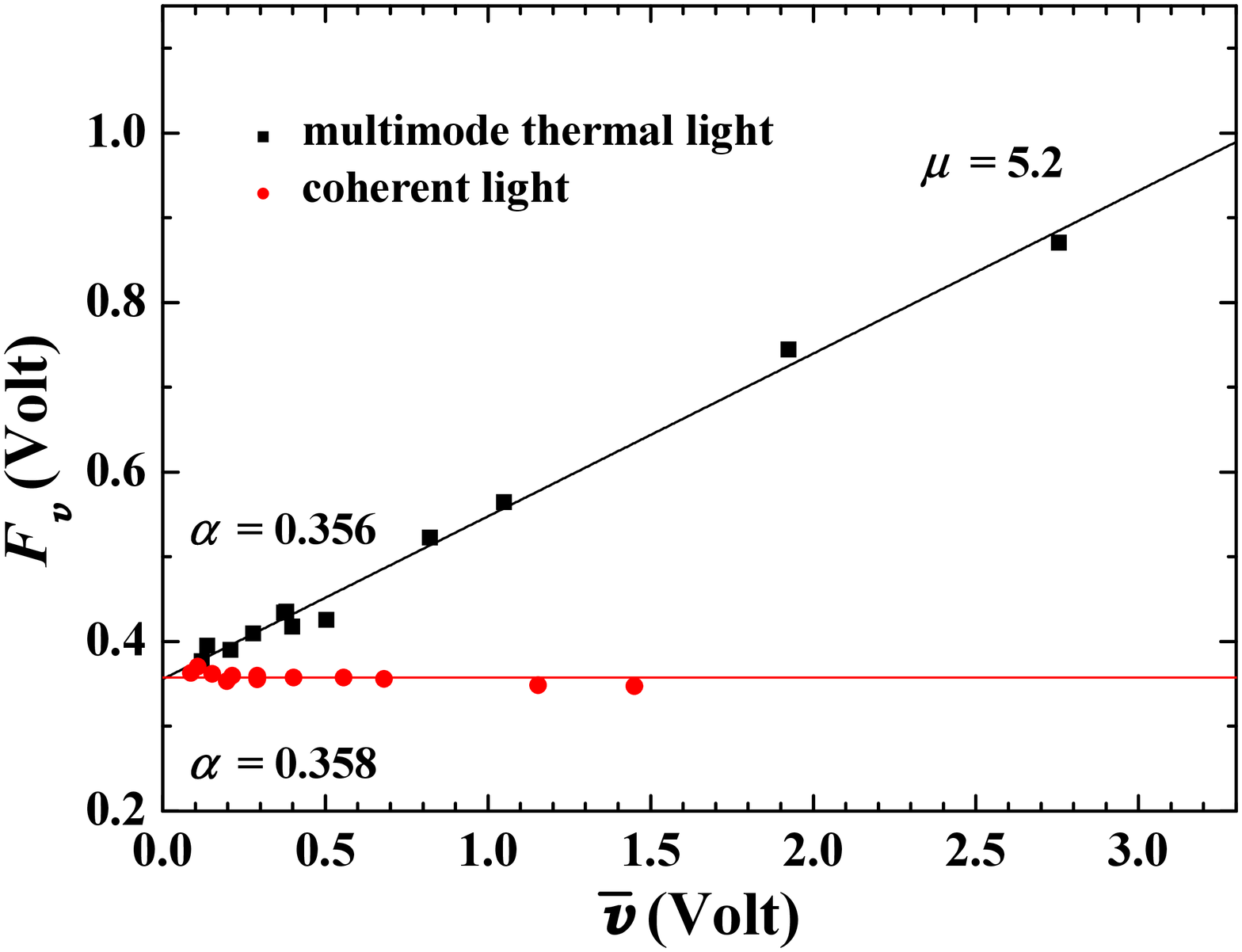}}
\caption{Plot of the Fano factor, $F_v$, for the output voltage as a function of the mean voltage $\bar{v}$ for different light states: pulsed coherent field at 523~nm (dots) and pulsed multimode thermal field at 420~nm (squares).}\label{f:fanoEXP}
\end{center}
\end{figure}

First of all we consider the measurements performed on ps-pulsed coherent light at 523~nm obtained as above (see Fig.~\ref{f:setup} ($a$)). For a coherent field we expect a Poissonian photon-number distribution
\begin{equation}
 P_{ph}(n)  = \frac{\bar{n}^n}{n!}\exp(-\bar{n})\ ,\label{eq:poiss}
\end{equation}
for which mean value and variance are $\bar{n}$. Note also that the photoelectron distribution obtained by applying Eq.~(\ref{eq:phel}) to Eq.~(\ref{eq:poiss}) remains Poissonian, so that $P_{el}(m)  = P_{ph}(n)$ with the substitution $\bar{n}\rightarrow \bar{m}$. The Fano factor for the final output voltage becomes
\begin{equation}
 F_v  = \alpha \ ,\label{eq:fanoCOH}
\end{equation}
independent of the mean value $\bar{v}$. The experimental results obtained with the PMT are shown in Fig.~\ref{f:fanoEXP} (full circles). The linear fit gives $\alpha = (0.358\pm 0.002)$~V.

The second kind of measurements is on the multimode thermal light given, as sketched in Fig.~\ref{f:setup} ($b$), by a blue portion (420~nm) of the down conversion parametric fluorescence produced by a type-I BBO crystal pumped by the third harmonics pulses (349~nm) of the same Nd:YLF laser \cite{paleari2004,bondani2007}. The expected photon-number distribution is given by the convolution of $\mu$ independent thermal modes \cite{mandel1995}
\begin{equation}
 P_{ph}(n)  = \frac{\left(n +\mu-1\right)!}
 {n!\left(\mu - 1 \right)! \left(\bar{n}/\mu+1 \right)^{\mu} \left(\mu/\bar{n}+1 \right)^{n}}\ , \label{eq:multit}
\end{equation}
for which the mean value is $\bar{n}$ and the variance is $\sigma^{(2)}_{ph}(n)= \bar{n}\left(\bar{n}/\mu+1 \right)$. As in the case of coherent light, the photoelectron distribution obtained by applying Eq.~(\ref{eq:phel}) to Eq.~(\ref{eq:multit}) remains multimode thermal and again $P_{el}(m)  = P_{ph}(n)$. The Fano factor for the final output voltage turns out to be
\begin{equation}
 F_v  = \frac{\bar{v}}{\mu}+\alpha\ ,\label{eq:fanoTHER}
\end{equation}
that is linear in the mean value. The experimental results obtained with the PMT are shown in Fig.~\ref{f:fanoEXP} (squares). Their fit to Eq.~(\ref{eq:fanoTHER}) gives $\mu = 5.2\pm 0.1$ and $\alpha = (0.356\pm 0.006)$~V. Note that, as these measurements were made in the same experimental conditions as those on the coherent field, the two values obtained for $\alpha$ are very similar to each other.

\begin{figure}
\begin{center}
\resizebox{0.6\columnwidth}{!}{%
 \includegraphics[width=0.4\textwidth,angle=270]{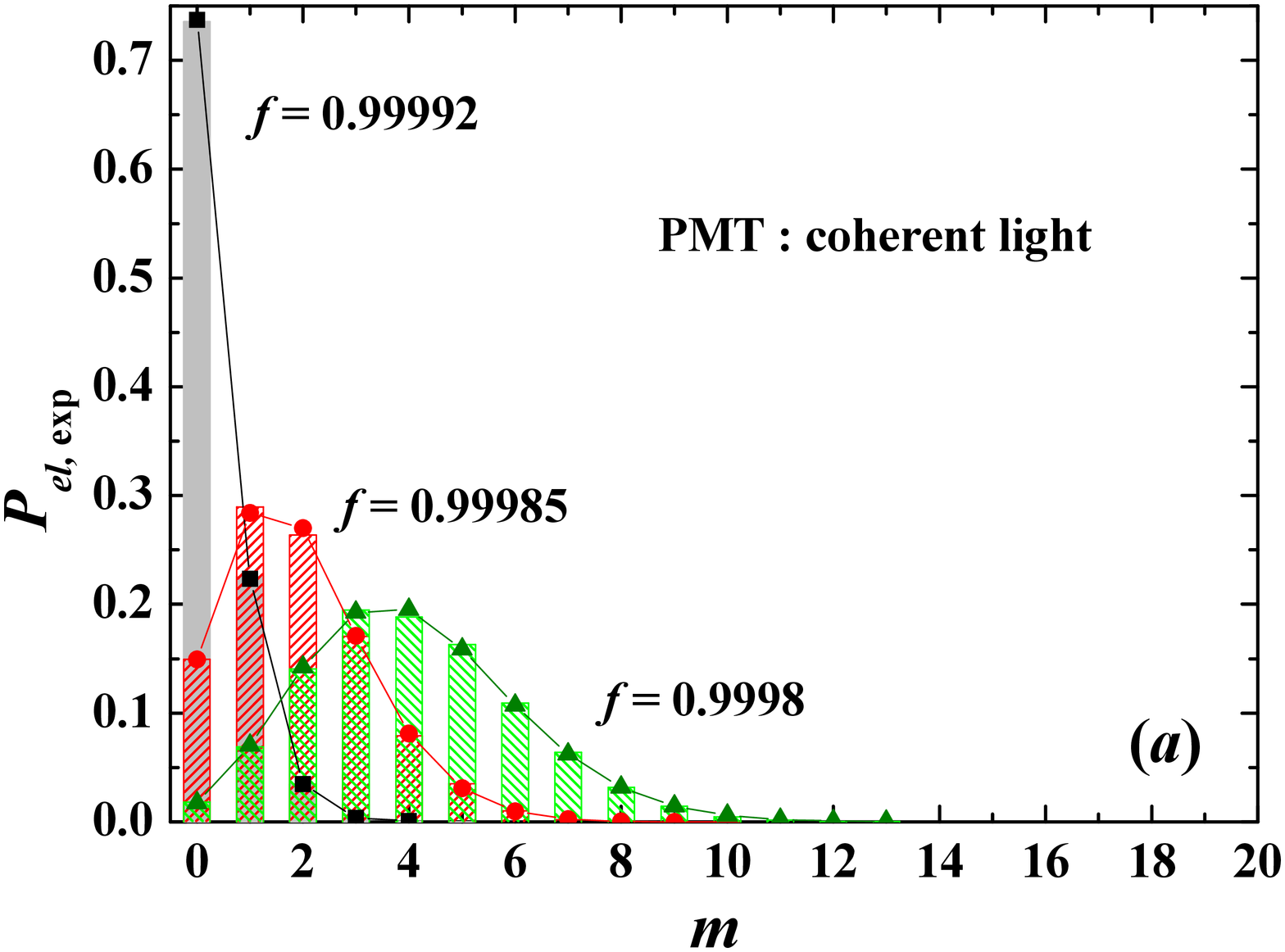}}
 \resizebox{0.6\columnwidth}{!}{%
 \includegraphics[width=0.4\textwidth,angle=270]{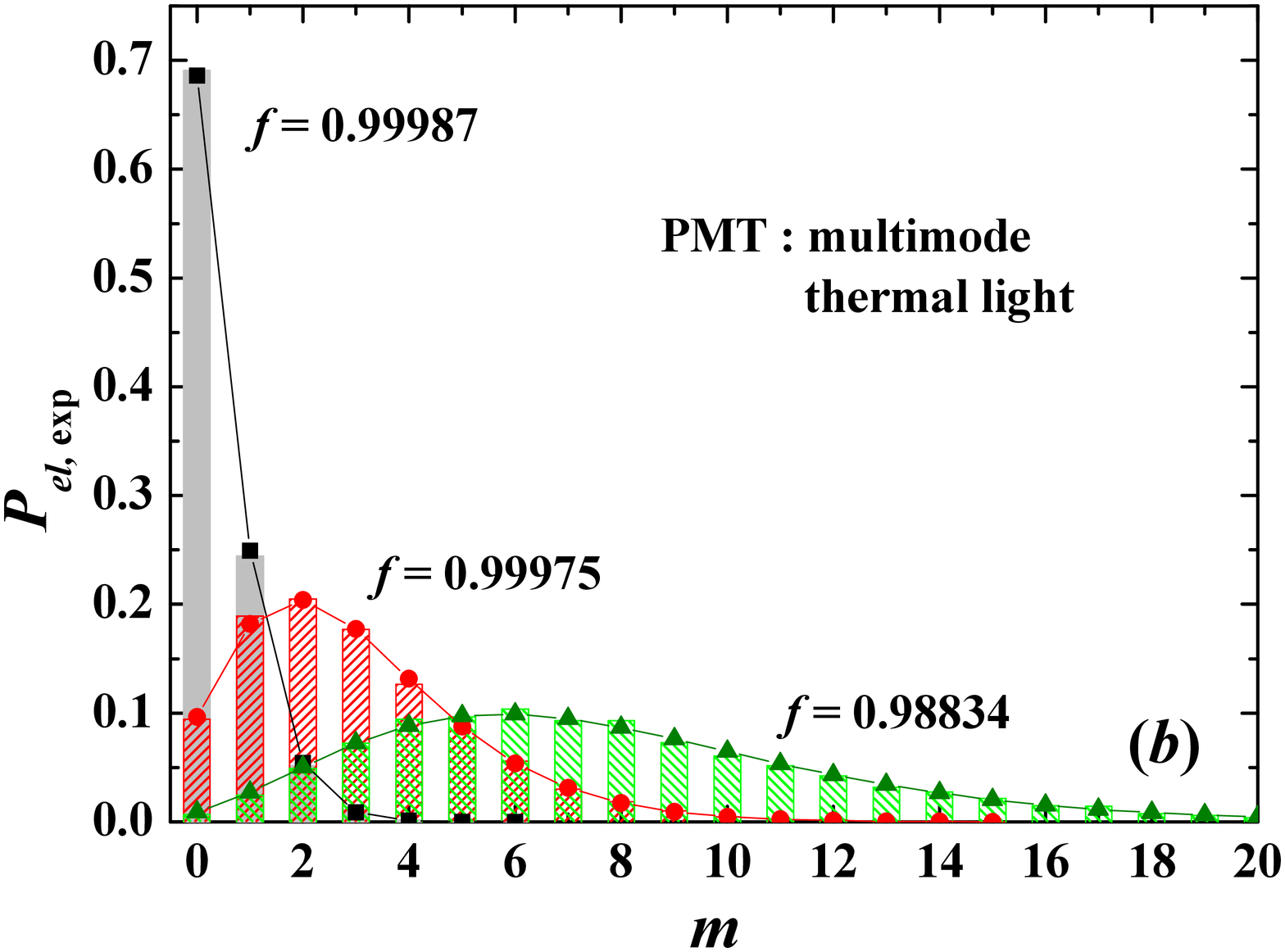}}
\caption{Reconstructed photoelectron distributions, $P_{el,\mathrm{exp}}$, for coherent light ($a$) and for multimode thermal light ($b$) reconstructed from some of the data sets used to obtain the calibration (bars). Symbols: theoretical curves, $P_{el}$, from  Eqs.~(\ref{eq:fanoCOH}) and (\ref{eq:fanoTHER}).} \label{f:reconstr}
\end{center}
\end{figure}

We now use the fitting parameters to recover the photoelectron statistics. We divide the voltage output by the value of $\alpha$ to convert it to photoelectrons and then re-bin the obtained distributions in unitary bins. In Fig.~\ref{f:reconstr} we show the photoelectron distributions, $P_{el,\mathrm{exp}}$ for coherent light ($a$) and for multimode thermal light ($b$) reconstructed from some of the data sets used to obtain the calibration. The dots in the figure are the theoretical curves, $P_{el}$ (Poissonian and multimode thermal, respectively), evaluated at the measured mean values, by using the parameters, $\alpha$ or $\alpha$ and $\mu$, obtained from the fits to Eqs.~(\ref{eq:fanoCOH}) and (\ref{eq:fanoTHER}), respectively.

In the figures the values of the fidelity \cite{jozsa1994} of the reconstructed distributions,
\begin{equation}
 f  = \sum_{m=1}^{\infty}\sqrt{P_{el,\mathrm{exp}}(m)P_{el}(m)} \ ,\label{eq:fidelity}
\end{equation}
are also displayed indicating the good quality of the reconstruction.

\begin{figure}
\begin{center}
\resizebox{0.6\columnwidth}{!}{%
 \includegraphics[width=0.4\textwidth,angle=270]{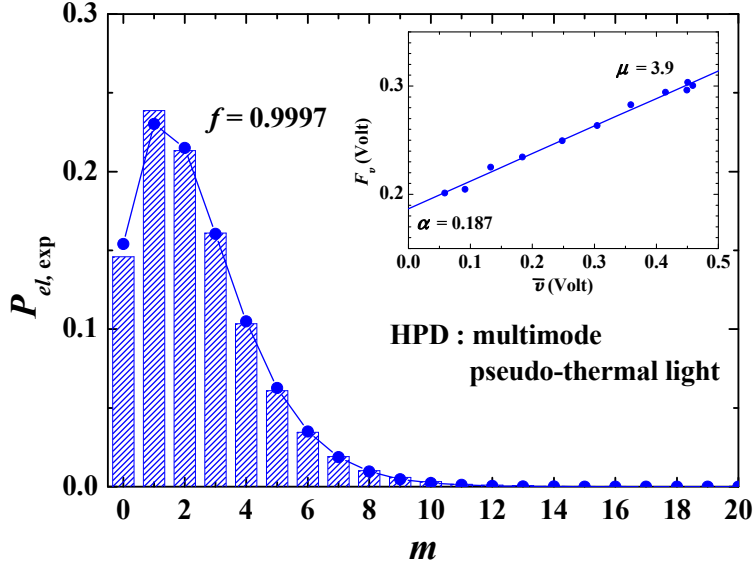}}
\caption{Reconstructed photoelectron distribution, $P_{el,\mathrm{exp}}$, for pulsed multimode pseudo-thermal light at 523~nm (bars). The displayed data set is one of those used to obtain the calibration. Symbols: theoretical curve, $P_{el}$, calculated from Eq.~(\ref{eq:fanoTHER}) with the number of modes ($\mu=3.9$) evaluated from the fit of the Fano factor (see Inset). Inset: plot of the Fano factor, $F_v$, for the output voltage as a function of the mean voltage $\bar{v}$.}\label{f:recHPD}
\end{center}
\end{figure}

The same kind of measurements were repeated by using the HPD. To optimize the detection efficiency we exploited the light produced by the second harmonics of the Nd:YLF laser. According to the setup depicted in Fig.~\ref{f:setup} (c), the laser light was passed through a rotating ground glass diffuser to obtain a pseudo-thermal speckle pattern \cite{arecchi1965}. The light, selected with a pin-hole so as to transmit a number of coherence areas, was then delivered to the detector by a multimode fiber. The photon-number statistics of the detected field follows the multimode thermal statistics in Eq.~(\ref{eq:multit}). In Fig.~\ref{f:recHPD} we plot the reconstructed photoelectron distribution for the multimode pseudo-thermal field along with the theoretical curve evaluated from the fit of the Fano factor, $\alpha=(0.187\pm 0.002)$~V and $\mu=3.9\pm 0.1$ (see Inset of the Figure). Again the fidelity of the reconstruction is very good.

\section{Conclusions}

We have demonstrated that it is possible to implement a self-consistent procedure to recover the distribution of photoelectrons that avoids calibration of the photodetectors. The procedure employs the evaluation of the Fano factor at different values of the mean photon numbers of the field to be characterized. The procedure has been satisfactorily tested on coherent and multimode thermal light. The next step will be to investigate less trivial optical states, such as those generated by the mixing of different fields \cite{zambra2007}, in order to extend the validity of the method.

\section*{Aknowledgements}

The Authors thank Sergio Cova (Politecnico di Milano) for fruitful discussions.\\
Present address of A. Agliati is: Quanta System, Solbiate Olona (VA), Italy.

\end{document}